\documentclass[conference]{IEEEtran}
\ifCLASSINFOpdf
\else
\fi
\usepackage{algorithm}
\usepackage{algpseudocode}
\usepackage{epsfig,makeidx,color}
\usepackage{amsmath,amssymb,bbm}
\usepackage{cite,graphicx,lipsum}
\usepackage{tabularx} 
\usepackage{float}    
\usepackage{enumerate}
\usepackage{booktabs}
\usepackage[switch,pagewise]{lineno}
\usepackage[bookmarks=false]{hyperref}
\usepackage[caption=false]{subfig}
\usepackage{tablefootnote}
\newcommand\tinyEarth{\vcenter{\hbox{\scalebox{0.5}{$\oplus$}}}}

\DeclareMathOperator{\asin}{asin}

\DeclareMathOperator{\atantwo}{atan2}

\title{On Delay Performance in Mega Satellite Networks with Inter-Satellite Links}

\author{Kosta {Dakic}$^\dagger$, Chiu Chun {Chan}$^\ddagger$, Bassel {Al Homssi}$^\star$,\\ Kandeepan Sithamparanathan$^\dagger$, and Akram {Al-Hourani}$^\dagger$\\
	$^\dagger$ School of Engineering, RMIT University, Melbourne, Australia\\
    $^\ddagger$ School of Engineering, Australian National University, Canberra, Australia\\
    $^\star$ School of Engineering \& Information Technology, UNSW Canberra, Canberra, Australia\\
Emails: kosta.dakic@ieee.org, akram.hourani@rmit.edu.au
}

\begin{document}
\maketitle

\begin{abstract}
Utilizing Low Earth Orbit (LEO) satellite networks equipped with Inter-Satellite Links (ISL) is envisioned to provide lower delay compared to traditional optical networks. However, LEO satellites have constrained energy resources as they rely on solar energy in their operations. Thus requiring special consideration when designing network topologies that do not only have low-delay link paths but also low-power consumption. In this paper, we study different satellite constellation types and network typologies and propose a novel power-efficient topology. As such, we compare three common satellite architectures, namely; (i) the theoretical random constellation, the widely deployed (ii) Walker-Delta, and (iii) Walker-Star constellations. The comparison is performed based on both the power efficiency and end-to-end delay. The results show that the proposed algorithm outperforms long-haul ISL paths in terms of energy efficiency with only a slight hit to delay performance relative to the conventional ISL topology. 
\end{abstract}

\begin{figure*}[ht]
    \normalsize
	\centering
   	\includegraphics[width=\textwidth]{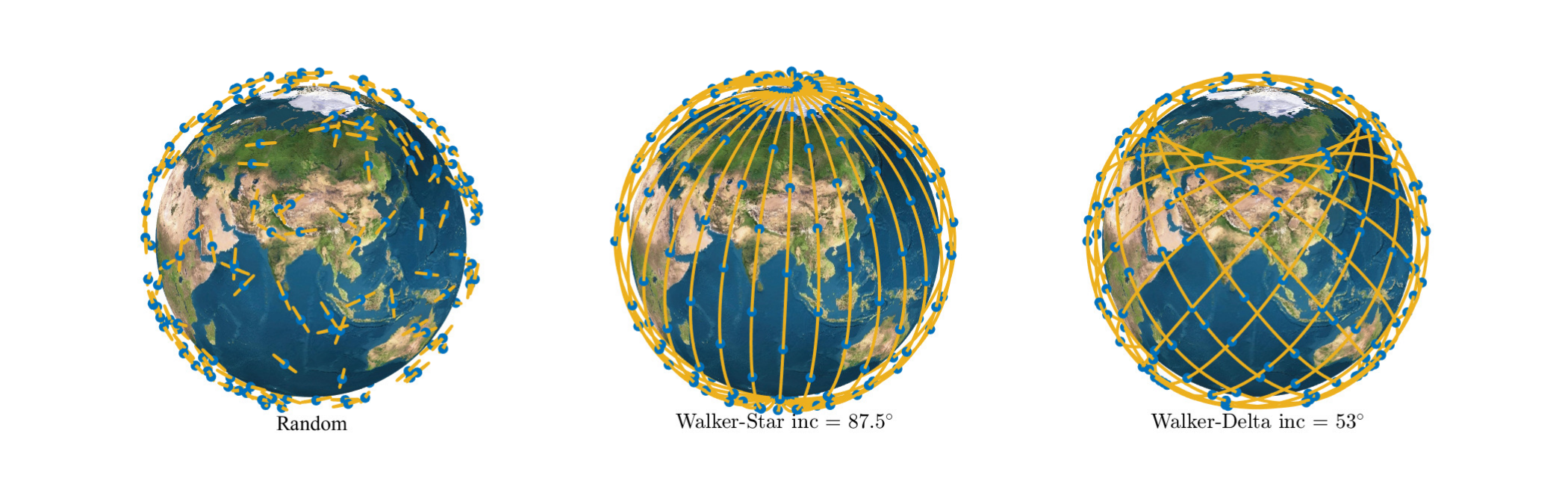}
	\caption{Different constellations evaluated in the study: random constellation (left), the Walker-Star with 16 orbital planes (middle), and the Walker-Delta with 16 orbital planes (right). The orange lines represent the satellite orbits, and satellites are illustrated as blue dots. The constellation size is 200 in this illustration.}
	\label{Fig: Constellationanditorbit}
\end{figure*}

\begin{figure}[!t]
    \normalsize
	\centering
   	\includegraphics[width=\linewidth]{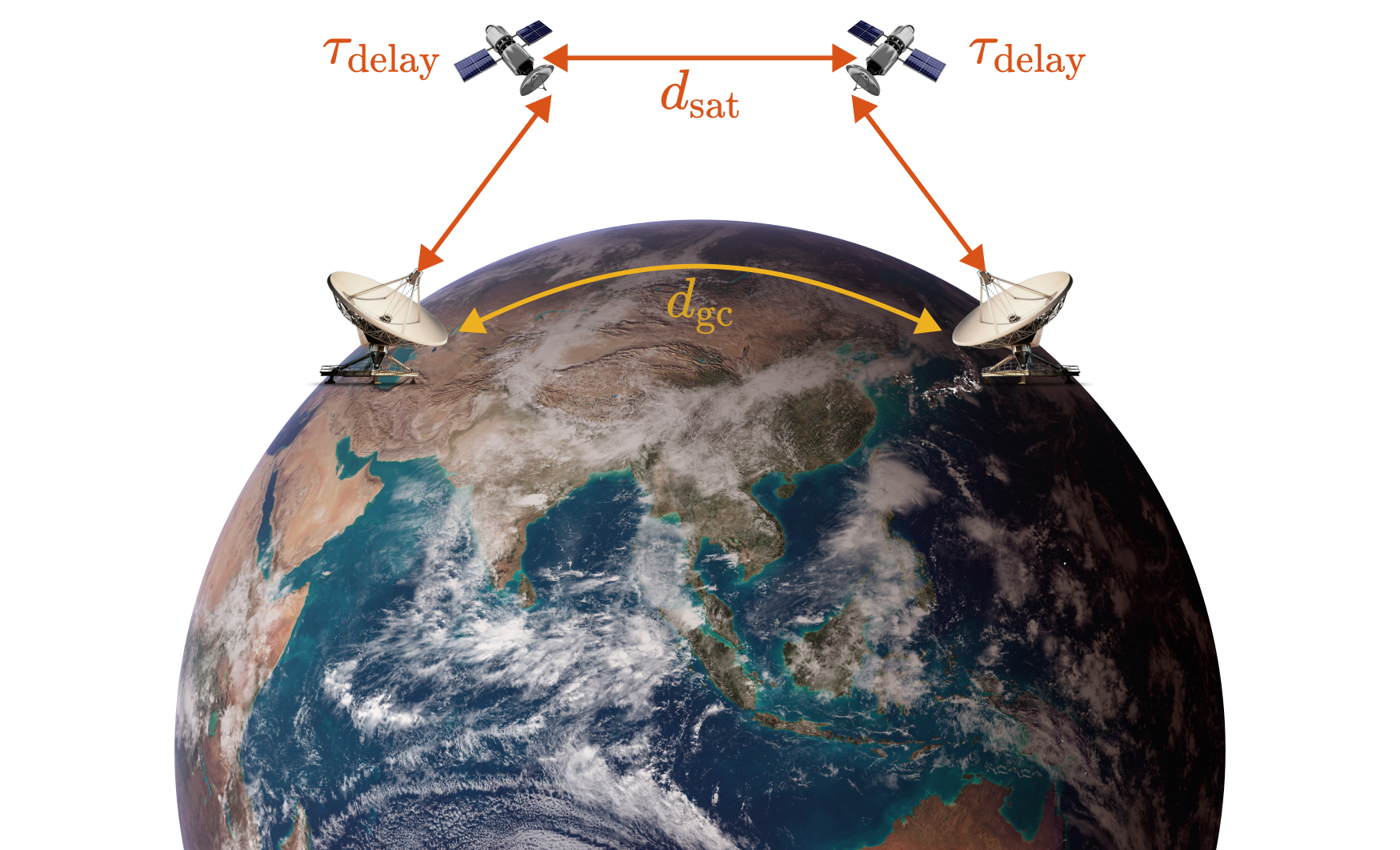}
	\caption{An illustration of the satellite distance and the great circle distance between two ground stations.}
	\label{FigISLvGround}
\end{figure}

\section{Introduction}\label{Sec_Intro}
\IEEEPARstart
The deployment of Low Earth Orbit (LEO) constellation networks is at an accelerating pace aiming to provide global connectivity. Many satellite projects, such as OneWeb, SpaceX's Starlink, and Amazon's Kupier~\cite{osoro2021techno} are currently been deployed with thousands of satellites into LEO orbit to provide seamless Internet coverage around the globe. These constellations will complement the existing terrestrial communications including both 5G and future 6G networks. 

One of the key enablers of such constellations envisioned inter-satellite connectivity. A connection between two satellites is referred to as Inter-Satellite Link (ISL) and is being widely anticipated (and demonstrated) in the upcoming LEO constellations. ISLs relay data directly between satellites, unlike current methods which depend on the large network of ground stations~\cite{BHmag}. ISLs could liberate constellations from the burden of establishing costly, and sometimes infeasible, ground station networks. An additional advantage is that the data carried by ISLs travel in free space and thus at the exact speed of light as opposed to conventional optical fiber networks. The average propagation speed in a typical single-mode optical fiber cables network is around $65-70\%$ the speed of light~\cite{udd2011fiber}. Hence, ISLs can potentially enable new low-delay applications such as remote control industry operations, cloud-controlled autonomous vehicles and farming, and telesurgery, in addition to enabling faster financial transactions. However, despite these advantages, satellite networks with ISLs face their own set of challenges, such as power constraints due to reliance on solar energy and the need to maintain reliable inter-satellite connections in a dynamic orbital environment. Consequently, the development of power-efficient and low-delay satellite constellation topologies that effectively leverage ISLs remains an active area of research and innovation.

Nevertheless, using satellites rather than optical fiber submarine cable possesses the potential to decrease the propagation delay by a few tens of milliseconds depending on the distance. Apart from delay reduction, the LEO satellite constellation can provide an access network for remote and rural communities, and to locations with extreme terrain such as mountains~\cite{BHmag}. This is also particularly important for industries that require real-time monitoring and control, such as manufacturing and logistics. Recent studies have also shown that LEO satellites with ISLs offer better resilience to cyber-attacks, making them a more secure option for data transfer in Industry 4.0 applications \cite{security,AlHomssi2022}. Furthermore, due to their low altitude, LEO satellites offer lower latency and higher bandwidth capacity compared to traditional geostationary satellites, enabling faster and more efficient communication services for users in remote areas~\cite{dakic2023spiking}. Recent studies have demonstrated the potential benefits of LEO satellite networks for improving connectivity in developing countries and bridging the digital divide~\cite{8002583}.


Simulations presented in~\cite{RN697} and in~\cite{markGround} illustrate the comparative delay advantages of ISL as a data relay technology versus optical cable. Additionally, authors in~\cite{RN714}, delve deeper into the idea of using optical satellite links rather than terrestrial fiber by developing a crossover function to optimize delay. The crossover function is a mathematical formula that determines the optimal point at which to switch from using a terrestrial fiber link to an optical satellite link. Authors in~\cite{27000} revealed the limitations of traditional network design approaches in the context of ISL-enable satellite communications and suggested the use of repetitive 3-satellite link patterns to address the temporal dynamics, achieving a higher efficiency than previous state-of-the-art methods. Nevertheless, more investigation is needed to better evaluate the benefits of optical satellite links for data relaying as well as to develop satellite-aware ISL topologies rather than just applying the shortest path ISLs. For example, a power-efficient ISL topology would take into consideration the power limitations of satellites due to their reliance on intermittent solar energy.



In this paper, we further analyze the prospect of using LEO satellite ISLs to relay data. The analysis is made through the simulation of different LEO satellite constellations (shown in Fig.~\ref{Fig: Constellationanditorbit}) where network topologies are proposed; (i) Nearest hop topology and (ii) Cutoff distance topology. We concentrate on the performance in terms of delay between the transmitting and receiving device because ISL is expected to facilitate lower delays, however, the performance of ISL-enabled networks needs to be carefully studied and assessed. An illustration of data communications with ISLs is shown in~\ref{FigISLvGround}. The contributions of this work are summarized as follows:
\begin{itemize}
    \item We compare the performance of three common satellite constellations, namely the theoretical random constellation, the widely deployed Walker-Delta, and the Walker-Star constellations in terms of end-to-end delay.
    \item We propose and evaluate two network topologies for LEO satellite constellations with ISLs: Nearest hop topology and Cutoff distance topology, focusing on their impact on delay performance comparing a theoretical great-circle optical fiber connection.
    \item We show that our proposed topologies achieve competitive delay performance relative to the conventional ISL topology, highlighting their potential for use in future LEO satellite networks.
\end{itemize}

\section{System Model}\label{Sec_Model}
\subsection{Geomtric model}
In this section, we introduce the three constellation models used for benchmarking the performance.
\subsubsection{Random}
The random distribution of satellites with random circular orbits. This is distribution is used in the simulations, as outlined in~\cite{RN706}, with the assumption that satellite collisions are disregarded. The left-hand side of Fig.~\ref{Fig: Constellationanditorbit} provides an illustration of the random constellation along with the common Walker constellations.

\subsubsection{Walker-Star Constellation} 
Satellite providers, such as OneWeb and Iridium use the Walker-Star constellation. The orbits in such constellation follow a near-polar configuration which has an inclination angle close to 90$^\circ$, this ensures global coverage, including the poles. However, this inherently results in an increased density of satellites with higher latitudes. The right ascension of the ascending node (RAAN) of the orbital planes in a Walker-Star configuration is spread across $0$ to $\pi$, unlike the Walker-Delta constellation which uses $0$ to $2\pi$. The middle of Fig.~\ref{Fig: Constellationanditorbit} depicts a typical Walker-Star constellation with 200 satellites along with their orbital planes.

\subsubsection{Walker-Delta Constellation} 
The Walker-Delta constellation, employed in satellite networks such as Kupier and Starlink~\cite{SpaceXFCC,KupierFCC}, reduces inter-satellite distance variations. These networks are currently considering inter-satellite links (ISLs) to support end-to-end communication without the need for a large network of terrestrial gateways. In the Walker-Delta configuration, the orbital planes are equally spaced and rotate around the Earth's axis of rotation, with a RAAN of $\Omega\in{0,\frac{2\pi}{P},2\frac{2\pi}{P},\dots,(P-1)\frac{2\pi}{P}}$. The right-hand side of Fig.~\ref{Fig: Constellationanditorbit} displays the constellation and orbital planes for the Walker-Delta constellation.

\subsection{Footprint model}
\begin{figure}[!t]
{\centering
\includegraphics[width=\linewidth]{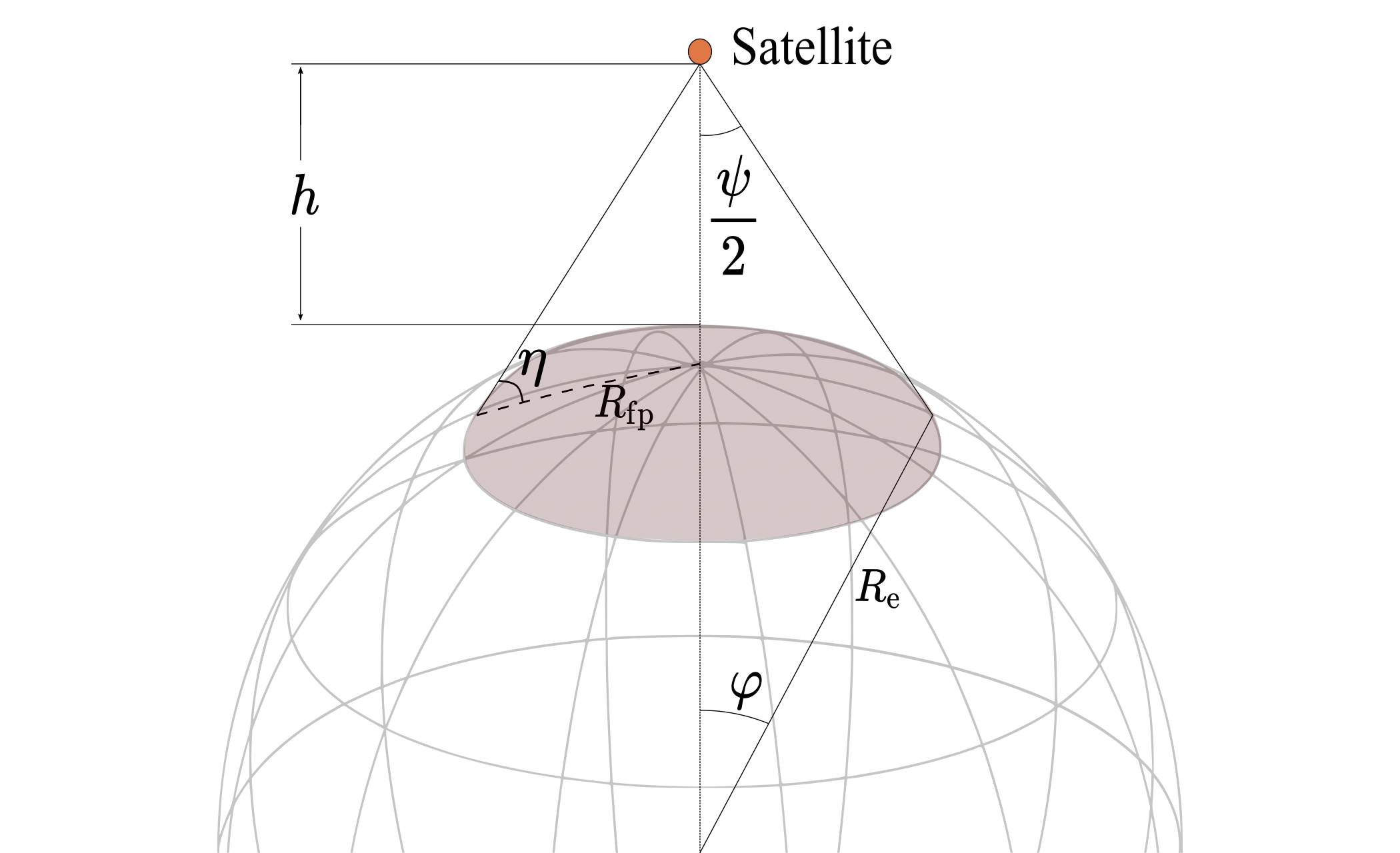}
\caption{The concept of Earth-centered zenith angle $\varphi$ and the satellite beamwidth $\psi$.}
\label{Fig_Geometry}}
\footnotesize
\end{figure}

In order to avoid terrestrial interference and heavy signal fading, the user-terminal only connects to satellites that have an elevation angle larger than a given threshold $\theta_\mathrm{min}$. One reasonable threshold is $25^\circ$ as FCC filings by Starlink~\cite{SpaceXFCC}. Thus, according to~\cite{9684552}, the effective footprint of a satellite is bounded by the minimum permissible elevation angle. In a practical sense, the footprint might be even smaller than this bound depending on the antenna beamwidth $\psi$. For the purpose of this study the footprint projection is assumed to be an ideal spherical cap bounded by an earth-centered zenith angle, denoted as $\varphi$, see Fig.~\ref{Fig_Geometry} for details. In order to calculate the beamwidth, the maximum slant distance between the satellite and the ground device needs to be calculated with the cosine rule as follows,
\begin{equation}
    a = R_\mathrm{e}\cos\left(\frac{\pi}{2}+\theta_\mathrm{min}\right)+\sqrt{R^2-R_\mathrm{e}\sin\left(\frac{\pi}{2}+\theta_\mathrm{min}\right)^2} ,
\end{equation}
where $R_\mathrm{e}$ is Earth's average radius, $R = R_\mathrm{e} + h$, and $h$ is the satellite altitude above the Earth's mean sea level. Thus, the maximum effective beamwidth can be again calculated with the cosine rule as follows~\cite{9684552},
\begin{equation}
    \psi = \mathrm{acos}\left(\frac{R_\mathrm{e}^2+R^2-a^2}{2Ra}\right) .
\end{equation}
Then the earth-centered zenith angle is calculated using the law of sines as follows~\cite{RN706},
\begin{equation}
    \varphi = \asin\left(\frac{1}{\alpha}\sin\frac{\psi}{2}\right) - \frac{\psi}{2} , 
\end{equation}
where $\alpha = R_\mathrm{e}/R$. Finally, the area of the spherical cap (footprint) of the beam is calculated as follows,
\begin{equation}
	A_\mathrm{fp} = 2\pi R_\mathrm{e}^2\left(1-\cos\varphi\right) ,
\end{equation}

The perimeter of a spherical cap can then be drawn on the earth's surface to define each satellite footprint, where if a device is located within the footprint, it is able to connect to the satellite. For defining the perimeter of the footprint, the latitude and longitude of the footprint boundary need to be calculated with the heading formulae~\cite{RN740} as follows,
\begin{equation}
	\phi_\mathrm{fp} = \asin\left(\sin \phi_\mathrm{sat} \cos \varphi + \cos \phi_\mathrm{sat} \sin \varphi \cos \theta\right) , 
\end{equation}
and the longitude,
\begin{multline}
	\rho_\mathrm{fp} = \rho_\mathrm{sat} + \atantwo(\sin \theta \sin \varphi \cos \phi_\mathrm{sat} , \\	 \cos \varphi - \sin \phi_\mathrm{sat} \sin \phi_\mathrm{fp}) ,
\end{multline}
where $\theta$ is an array from 0 to $2\pi$ with 360 elements and $\phi_\mathrm{sat}$ and $\rho_\mathrm{sat}$ is the latitude and longitude of the satellite in radians. An illustration of the geometry of a LEO satellite is shown in Fig.~\ref{Fig_Geometry}.

\begin{figure}[!t]
    \normalsize
	\centering
   	\includegraphics[width=\linewidth]{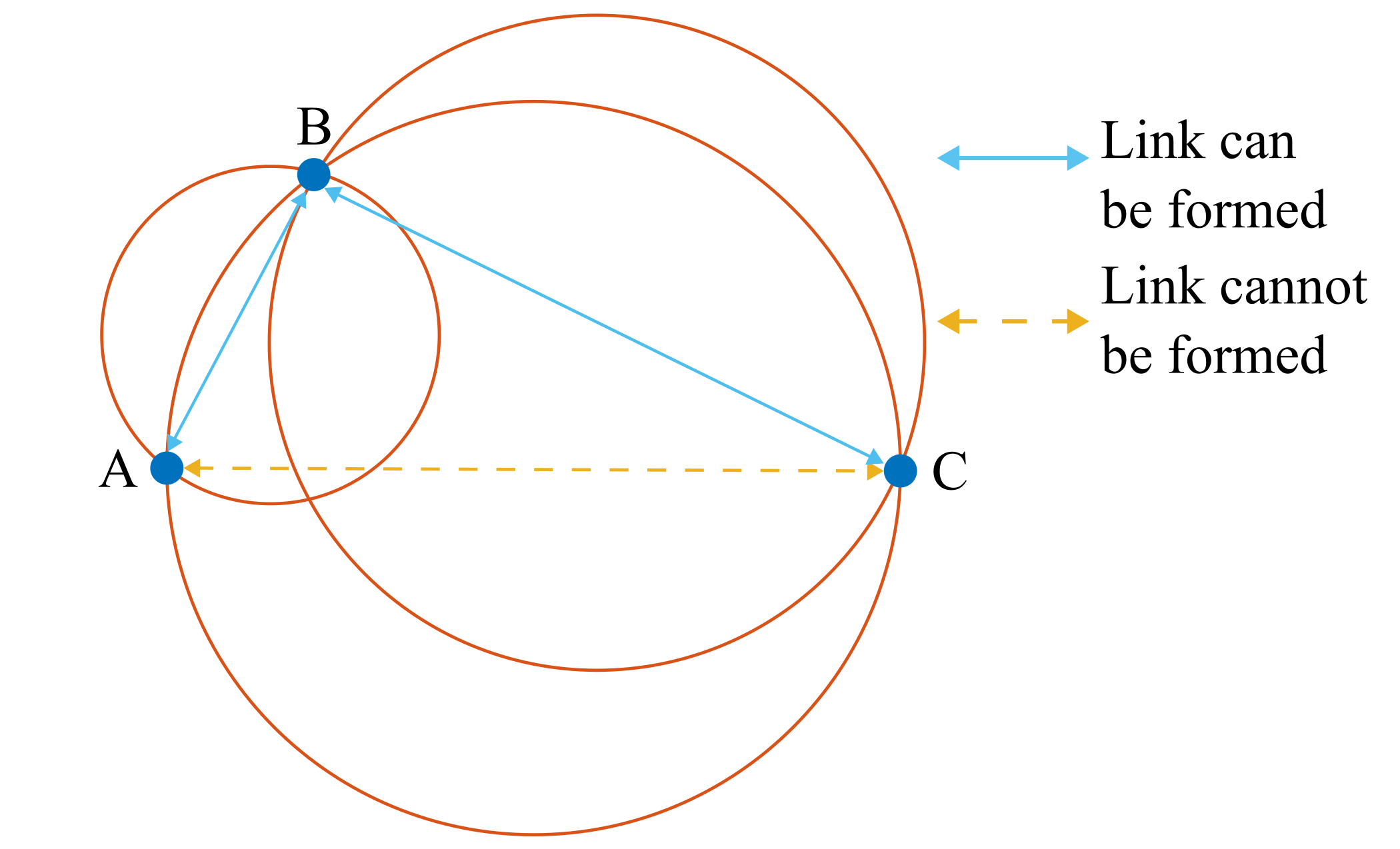}
	\caption{An illustration of the lowest energy next hop concept.}
	\label{ns_fig}
\end{figure}

\begin{figure*}[!t]
    \normalsize
	\centering
   	\includegraphics[width=\textwidth]{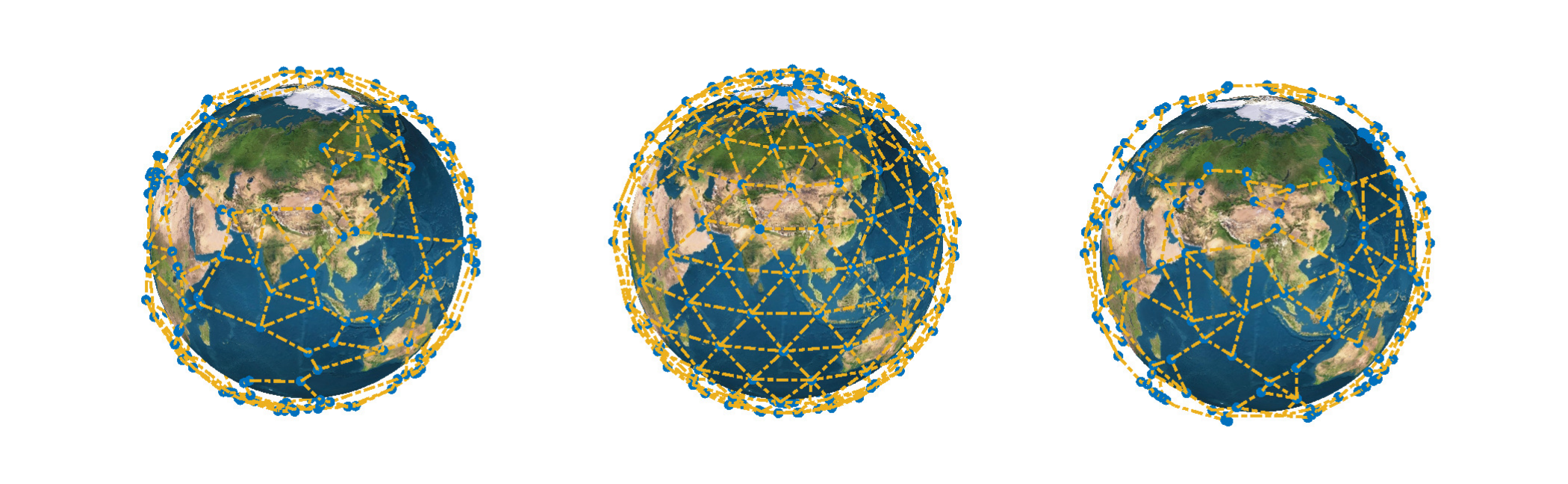}
	\caption{Different constellations with the nearest hop topology shown: random constellation (left), the Walker-Star (middle), and the Walker-Delta (right).}
	\label{Fig: Routing}
\end{figure*}

\subsection{Communication Delay}
In order to benchmark the satellite network, a direct point-to-point fiber link is assumed between the communicating ground terminals. The link thus follows the \textit{great-circle} which is the shortest path on a spherical surface. The delay is calculated as follows,
\begin{equation}
\tau_\mathrm{gc} = d_\mathrm{gc}/(c/n) 
\end{equation}\label{eq: Scurve_eq}
where $c$ is the speed of light, $n$ is the refractive index of the optical fiber cable, and $d_\mathrm{gc}$ is the great circle distance between the communicating ground terminals. When using the satellite network, the total delay for related to the distance of the sum of all hops distances plus the approximated processing delay and is formulated as follows,
\begin{equation}
    \tau_\mathrm{gc} = \left(d_\mathrm{sat}+d_\mathrm{tx\rightarrow sat}+d_\mathrm{sat \rightarrow rx}\right) /c + N_\mathrm{hops}\tau_\mathrm{process} ,
\end{equation}
where $d_\mathrm{sat}$ is the sum of all the ISL distances from the first satellite point to the last satellite point, $d_\mathrm{tx \rightarrow sat}$ is the distance from the transmitter to the first satellite, and $d_\mathrm{sat \rightarrow rx}$ is the distance from the last satellite in the path to the receiver. The processing delay is denoted as $\tau_\mathrm{process}$. 

\section{Topologies}
In this paper, we evaluate the end-to-end delay performance of different ISL-enabled constellations with two different network topologies. The topologies being utilized are (i) cutoff distance-based topology and (ii) nearest hop-based topology. For both topologies, after the links are constructed, Dijkstra's algorithm is used to calculate the lowest delay path. For Dijkstra's algorithm, the distance of each link is used for the link weight because it is directly proportional to the propagation delay.

\subsection{Nearest Hop-based Topology}
In a practical LEO satellite constellation, the number of available optical ISL ports (links) is limited due to various reasons including, cost, energy consumption, and satellite size. Therefore, ISLs links to neighboring satellites need to be optimized according to the given criteria. One way to form such links is to connect with the next satellites (next hop) that minimize the transmission energy. One method to establish the lowest energy next hops is to evaluate every satellite within a given vicinity and then pick up only the neighbors that if reached by a direct link would be more energy efficient. In geometric terms, this topology can be realized using the following steps:
\begin{enumerate}
    \item Draw a virtual sphere between the current satellite and the candidate satellite residing on the opposing side of the diameter segment.
    \item Create a link from the test node to the candidate only if there is no other candidate node in the sphere. 
    \item Repeat for every candidate satellite in the vicinity
\end{enumerate}
By referring to the illustration in Fig.~\ref{ns_fig}, the figure shows an ISL path from node A to node D. If we assume the transmission power is variable, the link from A $\rightarrow$ C cannot be made as link A $\rightarrow$ B is more efficient. The link between A and C cannot be made as $d_\mathrm{AC}^2 \ge d_\mathrm{AB}^2 + d_\mathrm{BC}^2$. The inverse-square relationship between distance and transmit power due to the free space path-loss (FSPL) exponent means that less transmit power is required if the signal travels A $\rightarrow$ B rather than directly from A $\rightarrow$ C. The FSPL is calculated as follows,
\begin{equation}
    l = \left[{\frac{4\pi d}{\lambda}}\right]^2 ,
\end{equation}

Then to calculate the total energy consumption of the system is formulated as follows,
\begin{equation} \label{pw}
     E = \alpha \sum_{i=0}^K d_{i}^2 + \left(E_\text{processing} \times K\right) ,
\end{equation}
where $K$ is the number of hops and $\alpha$ is the transmit power.

After all the ISLs are made, Dijkstra's algorithm~\cite{dijkstra1959note} is then used to calculate the shortest path between the transmitter and receiver using the satellite network. A figure showing the ISLs between each satellite using the nearest hop algorithm is shown in Fig.~\ref{Fig: Routing}. Additionally, the pseudocode for constructing the nearest hop topology is shown in~\ref{alg:cap}.

\begin{algorithm}
\caption{Nearest hop topology}\label{alg:cap}
\begin{algorithmic}
\Require $n > 0$ \Comment{Number of candidates}
\For{$i = 1~\text{to}~N$} \Comment{Loop through all satellites}
\State $Sat_\mathrm{candidate} \gets n$ number of closest satellites to satellite $Sat(i)$ 
\For{$p = 1~\text{to}~n$} \Comment{Check if a link can be made}
\State Draw sphere $Sat(i) \leftrightarrow Sat_\mathrm{candidate}(p)$ with Eq.(6) and (7)
\If{No other satellite in the sphere}
\State Connect $Sat(i) \leftrightarrow Sat_\mathrm{candidate}(p)$   
\EndIf
\EndFor
\EndFor
\State $\text{Tx}_\mathrm{sat} \gets$ closest satellite to $\text{Rx}_\mathrm{coord}(Lat,Lon)$ 
\State $\text{Rx}_\mathrm{sat} \gets$ closest satellite to $\text{Tx}_\mathrm{coord}(Lat,Lon)$ 
\State $path \gets \text{DIJKSTRA}\left(\text{Tx}_\mathrm{sat},\text{Rx}_\mathrm{sat}\right)$
\end{algorithmic}
\end{algorithm}

\subsection{Cutoff Distance -based Topology}
\begin{figure}[!ht]
    \normalsize
	\centering
   	\includegraphics[width=\linewidth]{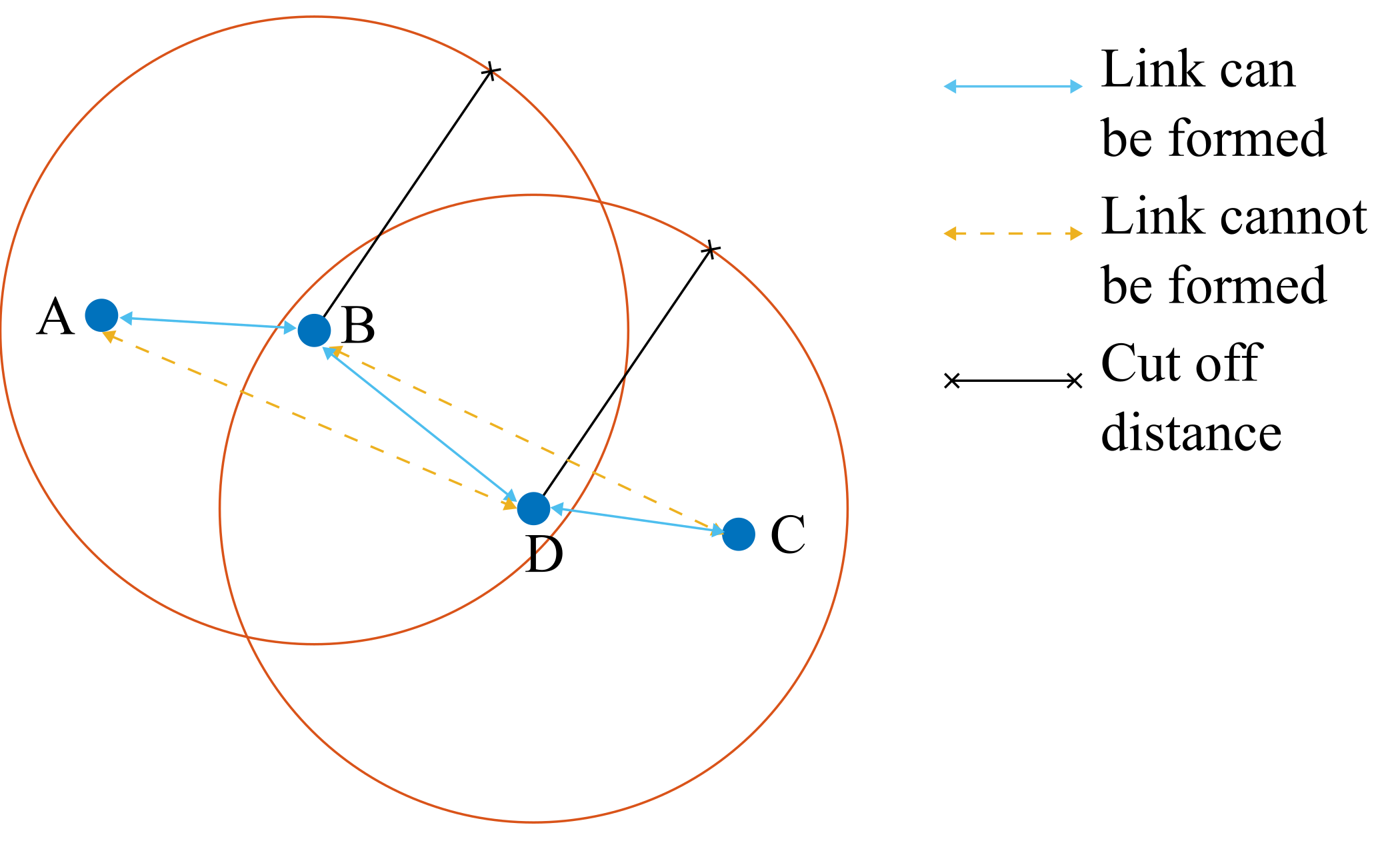}
	\caption{An illustration of the Cutoff concept.}
	\label{Fig_cutoff}
\end{figure}

This topology assumes that a satellite can link with any other satellite if it is closer that a given distance threshold. The practical sense of such a topology is that ISL links would have a maximum viable distance either limited by the link budget or by the occlusion caused by the Earth's curvature. For a generic case, we take the assumption that the links are limited by the Earth's curvature which imposes the upper bound distance of  feasible ISL links. Accordingly, the connection between the neighbor pairs is removed when the weight exceeds the maximum horizon range, denoted as $d_\mathrm{max}$. To be more accurate, the practical visibility constraint of ISL is limited by the troposphere which contains 99\% of the atmospheric water vapor and aerosols~\cite{troposphere}. The troposphere has an average height of around 18~km above the Earth’s surface so we use this as a threshold~\cite{troposphere}. The maximum ISL links distance is then calculated as follows,
\begin{equation}
    \centering
    d_\mathrm{max}= 2\sqrt{\left({R_\oplus}+h_\mathrm{s}\right)^2-\left({R_\oplus}+h_\mathrm{t}\right)^2} ,    \label{eq:horizoneq}
\end{equation}
where $h_\mathrm{s}$ is the height of the satellite and $h_\mathrm{t}$ is the average height of the troposphere. Dijkstra's algorithm~\cite{dijkstra1959note} is then utilized to find the shortest path in the topology. Note, the cutoff algorithm provides a lower bound for ISL delay, as this is the maximum distance at which ISLs could be formed. An illustration of how the links for the Cutoff routing concept are formed is shown in Fig.~\ref{Fig_cutoff}.

\begin{table}[!ht]
    \caption{Simulation Parameters.}
	\label{table1:SimulationParameter}
	\centering
    {\begin{tabular}{l l l}
    \toprule
    {\bf Parameters}  &{\bf Symbol}& {\bf Simulation value}\\
	\midrule
    Satellite constellation altitude& $h_\mathrm{s}$ & 550 [km]\\
    Troposphere altitude & $h_\mathrm{t}$ & 18 [km] \\
    Number of satellites & $N$ & [250 500 1000] \\
    Orbital planes & $P$ & 12\\
    Earth's radius & $R_{\tinyEarth}$ & 6371~[km]\\
    Satellite processing delay & $\tau_\mathrm{process}$ & 5.6~[$\mu$s] \\
   \bottomrule
	\end{tabular}}
	\begin{tabularx}{\textwidth}{l l}
	\end{tabularx}
\end{table}

\section{Results}
In this section, we present the performance results compared to torrential optical fiber connections between two arbitrary locations on Earth. Additionally, we explore three different LEO satellite constellations, (i) random, (ii) Walker-Star, and (iii) Walker-Delta. Note, we explore the performance on the random constellation as a baseline because it has been shown to be analytically tractable from coverage handover problems~\cite{ChiuConf, akramoptimalcoverage, Hybrid}.

For calculating the processing delay, we assume a processing clock operating at $533$~MHz, which is chosen as the Zynq UltraScale+ RFSoC~\cite{Zynq} as an example from a real-time single-chip radio platform. We also assume 3000 instructions to decide the ISL for the next satellite, the low number of instructions is due to the orbits being deterministic, thus all the topologies can be calculated apriori and the satellites would just need to use a look-up table to calculate the next link. The processing delay can then be calculated as $D_\mathrm{p} = \text{Number of Instructions} / {CLK}$. As the processing delay within satellites is likely to be fixed and similar to each other, we assumed the same processing delay for all satellites. Therefore, the processing delay is multiplied by the number of satellite hops $K$.

\subsection{Distance vs. Delay}
When comparing the delay of LEO satellite networks for data communication and the great circle optical fiber path, we normalize the delay to the great circle optical fiber path. The $\text{Improvement} = {D_\text{Optical fiber}}/{D_\text{Satellite}} - 1
$, where $D$ is the delay. As such, the improvement relative to the great circle optical fiber path is plotted relative to the great circle distance.

\begin{figure}[!t]
    \normalsize
	\centering
   	\includegraphics[width=\linewidth]{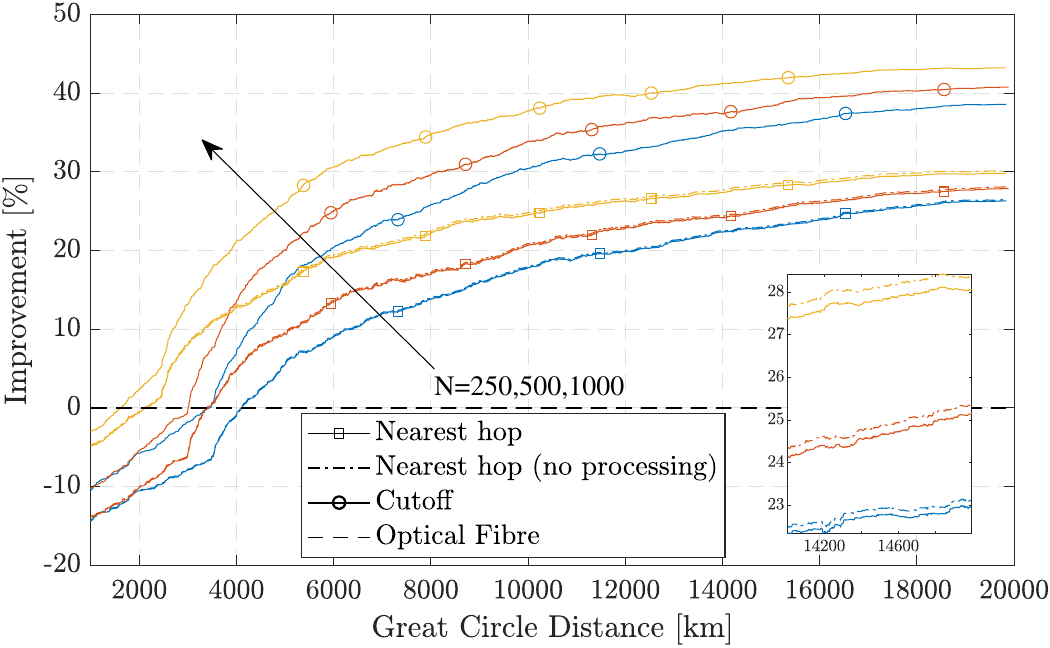}
	\caption{Delay plotted against the great circle distance between the transmitter and receiver, where the ISL delay is benchmarked against conventional optical fiber cable between the transmitter and receiver. }
	\label{Figdelay}
\end{figure}

The great circle distance is plotted against the delay in Fig.~\ref{Figdelay}, where we normalize the delay to the delay achieved by using an optical fiber path. The plot then shows a trend as the number of orbiting satellites increases, the performance also improves. The improvement is due to the greater coverage of satellites around the globe so the probability of a more efficient path existing also increases. Additionally in the same plot, the performance of the nearest hop topology is compared when processing is included in the calculation as well as assumed negligible. The plot shows the performance decreases very slightly due to the high clock rate of the hardware considered in this work~\cite{Zynq}. The performance of cutoff topology with and without processing was also considered, however, due to the smaller hop count relative to the nearest hop topology, the effect is negligible thus it is not shown on the plot. Note, the random constellation is used as a baseline example.

\begin{figure}[!t]
    \normalsize
	\centering
   	\includegraphics[width=\linewidth]{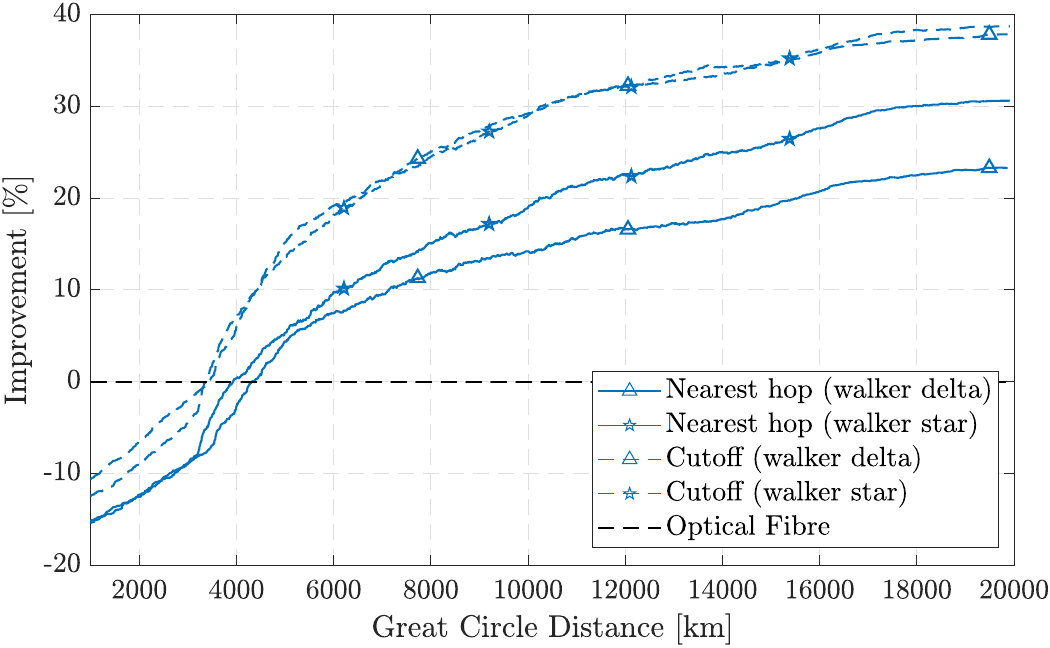}
	\caption{Delay plotted against the great circle distance between the transmitter and receiver. The ISL delay is benchmarked against conventional optical fiber cable between the transmitter and receiver with $N=250$. }
	\label{Figdelay_walker}
\end{figure}

In Fig.~\ref{Figdelay_walker} the performance of the delay against distance for LEO walker constellations, which are also normalized a great circle optical fiber path. From the plot, it can be seen that like with the random constellation, the performance in terms of delay increases as the number of satellites increases. Also, the Walker-Star seems to outperform the Walker-Delta constellation when the nearest hop topology is used. Furthermore, the performance contrast between the Walker-Delta and Walker-Star is the greatest when the nearest hop algorithm is used and also increases as the distance between the transmitting device and the receiving device grows. The performance disparity between the walker constellations is much small when the cutoff topology is used. The Walker-Star constellation outperforms the Walker-Delta constellation in terms of delay as the Walker-Star has better coverage over the globe compared to the Walker-Delta and random locations for the ground-based transmitter and receiver are used. 

\begin{figure}[!t]
    \normalsize
	\centering
   	\includegraphics[width=\linewidth]{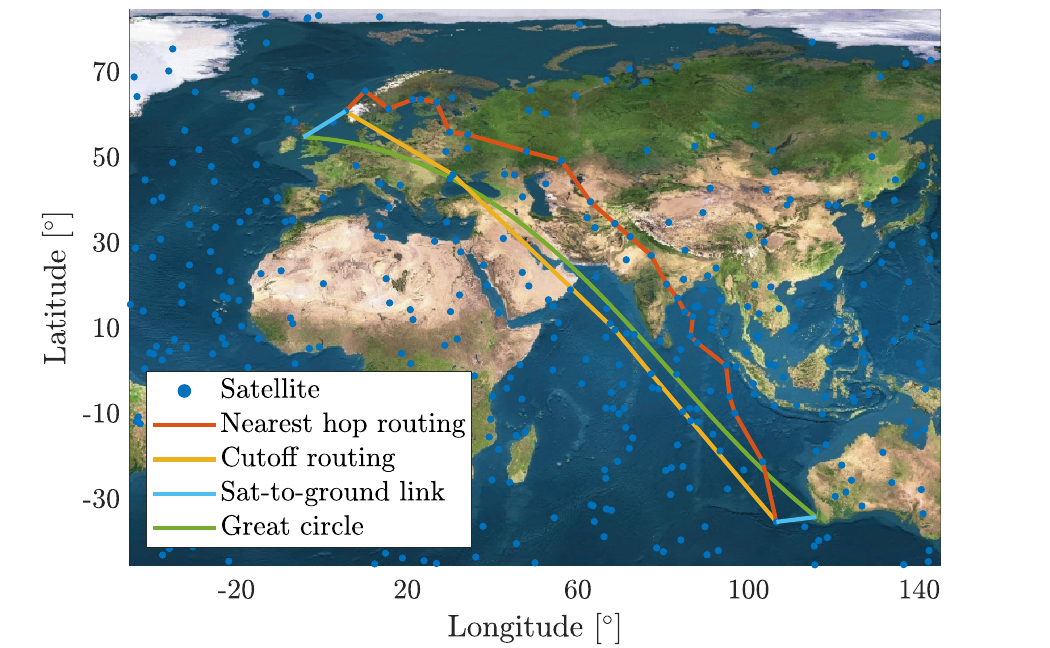}
	\caption{Simulation of a random constellation, where the transmitting device is in Perth, Australia, and the receiving device is in Brest, France. The shortest hop path is shown in orange while the cutoff topology ISL path is shown in yellow. }
	\label{Figpath}
\end{figure}

\subsection{Alternate Paths}
\begin{figure*}[!t]
    \normalsize
	\centering
   	\includegraphics[width=\textwidth]{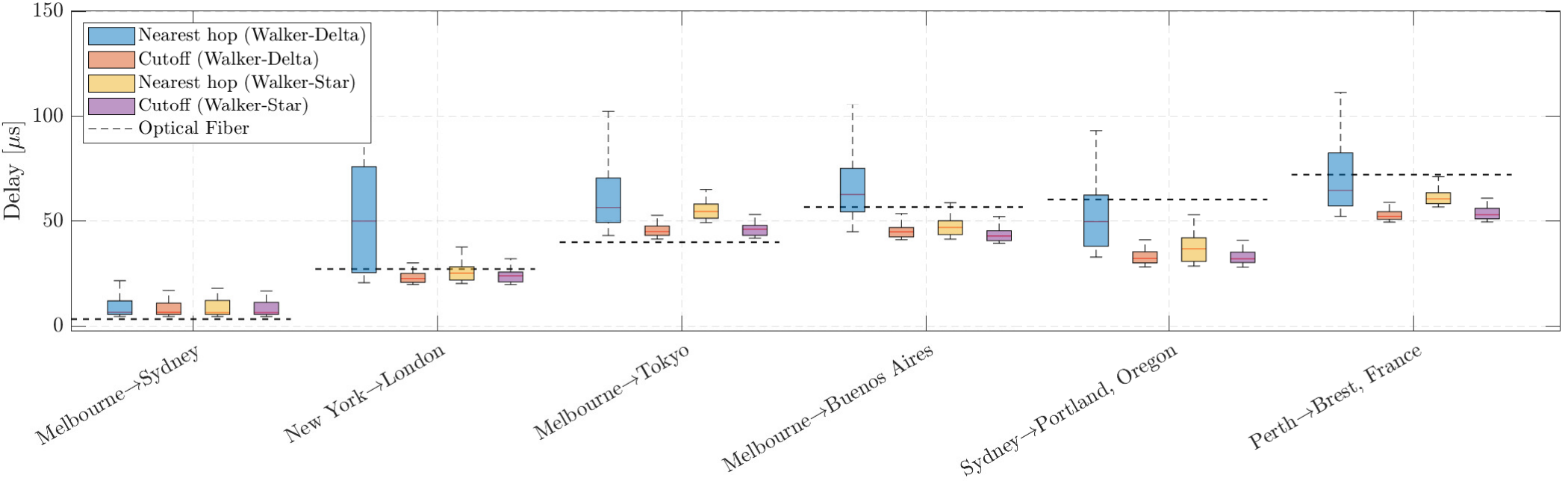}
	\caption{The delay between major cities with $N=1000$.}
	\label{Fig_Altpath}
\end{figure*}

To determine the robustness of each LEO constellation to link failures, we investigate how the link delay increases as we introduce alternate paths. We construct the best possible path and calculate the delay, then remove the links that correspond to the best path and form a new path. The delay of the new path is then calculated. The process is repeated until we have 10 distinct paths from the transmitting device to the receiving device. The results are shown in Fig.~\ref{Fig_Altpath} where the delay performance is shown for Walker-Delta as well as Walker-Star satellite constellations. In addition, the delay for using terrestrial optical fiber cable over the great circle distance between the cities is shown. An illustration showing the different types of links is shown in Fig.~\ref{Figpath}. The Walker-Delta constellation using the nearest hop topology has the greatest deviation between cities, particularly for the path between New York and London. Moreover, the deviation from using the cutoff algorithm is very low, thereby showing greater robustness to failed links between satellites. Finally, the performance of satellites for ISL paths compared to traditional fiber has a greater pay-off when the distance between the two locations increases such as in the delay between Perth and Brest.

\section{Conclusion}\label{Sec_Conc}
In this study, we analyzed ISL-enabled LEO satellite constellations as a low-delay alternative to traditional terrestrial optical fiber networks. We investigated three LEO constellations: Walker-Delta, Walker-Star, and random. Additionally, we explored two topologies, presenting the power-efficient nearest hop topology and comparing it to the delay-minimizing cutoff topology. Our results demonstrated that satellite networks improve delay performance compared to optical fiber connections as the transmitter-receiver distance increases. The proposed nearest hop topology maintains a better delay performance compared to the great-circle fiber path while also utilizing more energy-efficient ISLs. Future work will involve using machine learning to develop a topology for dynamically changing ISLs due to high traffic loads and link failures.


The Walker-Star constellation had a lower delay, but the Walker-Delta constellation was more power-efficient in terms of average ISL length with the nearest hop algorithm.  

\section{Acknowledgment}
The authors would like to acknowledge the discussions with Dr. Ben Allen and Mr. Ben Moores, also the partial funding by SmartSat CRC under the UK-Australia spacebridge program.
\bibliographystyle{IEEEtran}
\bibliography{main}
\end{document}